\documentclass[twocolumn,american,english]{revtex4}
\usepackage[T1]{fontenc}
\usepackage[latin9]{inputenc}
\usepackage{color}
\usepackage{float}
\usepackage{amsmath}
\usepackage{graphicx}

\makeatletter

\newcommand*\LyXZeroWidthSpace{\hspace{0pt}}
\providecommand{\tabularnewline}{\\}

\@ifundefined{textcolor}{}
{%
 \definecolor{BLACK}{gray}{0}
 \definecolor{WHITE}{gray}{1}
 \definecolor{RED}{rgb}{1,0,0}
 \definecolor{GREEN}{rgb}{0,1,0}
 \definecolor{BLUE}{rgb}{0,0,1}
 \definecolor{CYAN}{cmyk}{1,0,0,0}
 \definecolor{MAGENTA}{cmyk}{0,1,0,0}
 \definecolor{YELLOW}{cmyk}{0,0,1,0}
}

\makeatother

\usepackage{babel}
\begin{document}
\title{\textcolor{black}{Performance of the NOF-MP2 method in hydrogen abstraction
reactions.}}
\author{Xabier Lopez$^{1,2}$, Mario Piris$^{1,2,3}$}
\address{\bigskip{}
$^{1}$Kimika Fakultatea, Euskal Herriko Unibertsitatea (UPV/EHU),
P.K. 1072, 20080 Donostia, Euskadi (Spain);\\
 $^{2}$Donostia International Physics Center (DIPC), 20018 Donostia,
Euskadi (Spain);\\
 $^{3}$IKERBASQUE, Basque Foundation for Science, 48013 Bilbao, Euskadi
(Spain).}
\begin{abstract}
The recently proposed \textcolor{black}{natural orbital functional
second-order Møller\textendash Plesset (NOF-MP2) method} is capable
of achieving both dynamic and static correlation even for those systems
with significant multiconfigurational character. We test its reliability
to describe the electron correlation in radical formation reactions,
namely, in the homolytic X-H bond cleavage of LiH, BH, CH$_{4}$,
NH$_{3}$, H$_{2}$O and HF molecules. Our results are compared with
CASSCF and CASPT2 wavefunction calculations and the experimental data.
For a dataset of 20 organic molecules, the thermodynamics of C-H homolytic
bond cleavage, in which the C-H bond is broken in the presence of
different chemical environments, is presented. The radical stabilization
energies obtained for such general dataset are compared with the experimental
data. It is observed that NOF-MP2 is able to give a quantitative agreement
for dissociation energies, with a performance comparable to that of
the accurate CASPT2 method.
\end{abstract}
\maketitle

\section{Introduction}

Natural orbital functional (NOF) theory \citep{Piris2007} is being
configured as an alternative formalism to both DFT and wavefunction
methods, by describing the electronic structure in terms of the natural
orbitals (NOs) and their occupation numbers (ONs). Various functionals
have been developed in the last years, a comprehensive review can
be found in Refs. \citep{Piris2014a,Pernal2016}. Recently \citep{Piris2017},
a single-reference global method for the electron correlation was
introduced taking as reference the Slater determinant formed with
the NOs of an approximate NOF. In this approach, called \textcolor{black}{natural
orbital functional - second-order Møller\textendash Plesset (NOF-MP2)}
method, the total energy of an N-electron system can be attained by
the expression
\begin{equation}
E=\tilde{E}_{hf}+E^{corr}=\tilde{E}_{hf}+E^{dyn}+E^{sta}\label{Etotal}
\end{equation}

where $\tilde{E}_{hf}$ is the Hartree-Fock energy obtained with the
NOs, the dynamic energy ($E^{dyn}$) is derived from a modified MP2
perturbation theory, while the non-dynamic energy ($E^{sta}$) is
obtained from the static component of the employed NOF.

In fact, NOF theory is a particular case of the one-particle reduced
density matrix (1RDM) functional theory \citep{Gilbert1975,Levy1979,Valone1980},
in which the spectral decomposition of the 1RDM is assumed. In this
representation, restrictions on the ONs to the range $\left[0,1\right]$
represent the necessary and sufficient conditions for ensemble N-representability
of the 1RDM \citep{Coleman1963} under the Lowdin's normalization.
The exact functional in terms of the 1RDM has been an unattainable
goal so far, and we really work with approximations. Approximating
the energy functional implies that the functional N-representability
problem arises \citep{Piris2018}. To date, only NOFs proposed by
Piris and coworkers \citep{Piris2013b} rely on the reconstruction
of the two-particle reduced density matrix (2RDM) subject to ensemble
N-representability conditions.

The success of the NOF-MP2 method is determined by the NOs used to
generate the reference. The functional PNOF7s proved \citep{Piris2018b}
to be the functional of choice for the method. The \textquotedbl s\textquotedbl{}
emphasizes that this interacting-pair model takes into account only
the static correlation between pairs, and therefore avoids double
counting in the regions where the dynamic correlation predominates,
already in the NOF optimization. Moreover, \textcolor{black}{the correction
}$E^{dyn}$\textcolor{black}{{} is based on the orbital-invariant formulation
of the MP2 energy }\citep{Saebo2002}\textcolor{black}{.}

In the present paper, we analyze the performance of NOF-MP2 in the
description of X-H bond dissociations, important process in biological
\citep{Valko2006,Valko2007} and organic chemistry \citep{Breher2007}.
Firstly, we evaluate the dissociation energy for the X-H bonds in
LiH, BH, CH$_{4}$, NH$_{3}$, H$_{2}$O and HF molecules. Results
are compared to our previous calculations \citep{Lopez2015}, and
the experimental data.

The proper description of the X-H homolytic bond dissociation curves
is a fundamental step for the accurate characterization of the electronic
structure of these important species \citep{Basch1997,Coote2004,Temelso2006,Vandeputte2007}.
This requires the appropriate treatment of strong correlation effects
since a single Slater determinant wavefunction leads to incorrect
results. We need to include several determinants that lead to computationally
demanding methods. An alternative is the density functional theory
(DFT), however, it suffers from methodological problems to treat strong
electron correlation or near-degeneracy effects \citep{Shao2003,Krylov2006a,Ess2012}.
It is worth noting that cost-effective bond dissociation energies
can be obtained in the context of spin-dependent DFT, but at the price
of obtaining solutions with breaking symmetry \citep{Menon2007,Menon2008}.
Valence bond theory has also been used for this type of systems \citep{Lai2012}.

The formation of radicals by hydrogen abstraction is a fundamental
step to explain the oxidation of hydrocarbons \citep{Carstensen2007,Huynh2008,Huynh2010},
lipid-peroxidation \citep{Valko2007}, formation of reactive oxygen
species \citep{Stadtman2003}, Fenton chemistry \citep{Prousek2007}
and DNA damage \citep{Balasubramanian1998}. Due to this widespread
interest on the thermodynamic stability of organic radicals, we analyze
secondly the cleavage of the C-H bond in a dataset of 20 organic molecules,
previously designed in our group \citep{Lopez2015}. As a measure
of radical stability we employ the bond dissociation energy ($\mathrm{D_{e}}$)
which has often been used in the literature \citep{Vereecken2001,Blanksby2003}.
Based on $\mathrm{D_{e}}$, we estimate the radical stabilization
energy (RSE) for a variety of hydrogen abstraction reactions of the
type:
\begin{equation}
\mathrm{XH}+\mathrm{Y}^{.}\rightarrow\mathrm{X}^{.}+\mathrm{YH}\label{eq:isodesmicEq}
\end{equation}

RSE is equivalent to the difference in bond dissociation energies
of XC-H and Y-H species.

\section{Theory}

In this work, we address only singlet states, so we adopt the spin-restricted
theory in which a single set of orbitals is used for $\alpha$ and
$\beta$ spins. We shall use PNOF7s \citep{Piris2018b}, which is
a NOF based on the electron-pairing approach in NOF theory \citep{Piris2018a}.

Consider the orbital space $\Omega$ is divided into N/2 mutually
disjoint subspaces $\Omega{}_{g}$, so each orbital belongs only to
one subspace. Each subspace contains one orbital $g$ below the level
N/2, and $\mathrm{N}_{g}$ orbitals above it, which is reflected in
additional sum rules for the ONs,
\begin{equation}
{\displaystyle \sum_{p\in\Omega_{g}}}n_{p}=1,\quad g=1,2,\ldots,\mathrm{N}/2\label{sumrule_n}
\end{equation}
Taking into account the spin, each subspace contains only an electron
pair. The Lowdin's normalization condition is automatically fulfilled,
\begin{equation}
\begin{array}{c}
{\displaystyle 2\sum\limits _{p\in\Omega}n_{p}=2\sum_{g=1}^{\mathrm{N}/2}}{\displaystyle \sum_{p\in\Omega_{g}}}n_{p}=\mathrm{N}\end{array}
\end{equation}
Coupling each orbital $g$ below the N/2 level with only one orbital
above it ($\mathrm{N}_{g}=1$) leads to the orbital perfect-pairing
approach. In general, we fix $\mathrm{N_{g}}$ to the maximum allowed
value determined by the basis set used in calculations. It is important
to note that orbitals satisfying the pairing conditions (\ref{sumrule_n})
are not required to remain fixed throughout the orbital optimization
process \citep{Piris2009a}.

The energy of PNOF7s can be conveniently written as
\begin{equation}
\begin{array}{c}
E=\sum\limits _{g=1}^{\mathrm{N}/2}E_{g}+\sum\limits _{f\neq g}^{\mathrm{N}/2}E_{fg}\\
\\
E_{g}=\sum\limits _{p\in\Omega_{g}}n_{p}\left(2\mathcal{H}_{pp}+\mathcal{J}_{pp}\right)+\sum\limits _{p,q\in\Omega_{g},p\neq q}\Pi_{qp}^{g}\mathcal{L}_{pq}\\
\\
E_{fg}=\sum\limits _{q\in\Omega_{f}}\sum\limits _{p\in\Omega_{g}}\left[n_{q}n_{p}\left(2\mathcal{J}_{pq}-\mathcal{K}_{pq}\right)+\Pi_{qp}^{s}\mathcal{L}_{pq}\right]
\end{array}\label{EPNOF7s}
\end{equation}
where
\begin{equation}
\begin{array}{c}
\Pi_{qp}^{g}=\left\{ \begin{array}{cc}
-\sqrt{n_{q}n_{p}}\,, & p=g\textrm{ or }q=g\\
+\sqrt{n_{q}n_{p}}\,, & p,q>\mathrm{N}/2
\end{array}\right.\\
\\
\Pi_{qp}^{s}=-4n_{q}\left(1-n_{q}\right)n_{p}\left(1-n_{p}\right)\quad\;
\end{array}\label{Pi}
\end{equation}

$\mathcal{J}_{pq}$, $\mathcal{K}_{pq}$, and $\mathcal{L}_{pq}$
are the usual direct, exchange, and exchange-time-inversion two-electron
integrals. The first term of the energy in Eq. (\ref{EPNOF7s}) draws
the system as independent N/2 electron pairs, whereas the second term
contains the interactions between electrons belonging to different
pairs. PNOF7s provides the reference NOs to form $\tilde{E}_{hf}$
in the NOF-MP2 method, Eq. (\ref{Etotal}).

$E^{sta}$ is the sum of the static intrapair and interpair electron
correlation energies:
\begin{equation}
\begin{array}{c}
E^{sta}=\sum\limits _{g=1}^{\mathrm{N}/2}\sum\limits _{q\neq p}\sqrt{\Lambda_{q}\Lambda_{p}}\,\Pi_{qp}^{g}\mathcal{\,L}_{pq}+\sum\limits _{f\neq g}^{\mathrm{N}/2}\sum\limits _{p\in\Omega_{f}}\sum\limits _{q\in\Omega_{g}}\Pi_{qp}^{s}\mathcal{\,L}_{pq}\end{array}\label{Esta}
\end{equation}

where $\Lambda_{p}=1-\left|1-2n_{p}\right|$ is the amount of intra-pair
static correlation in each orbital as a function of its occupancy.

$E^{dyn}$ is obtained from the second-order correction $E^{\left(2\right)}$
of the MP2 method. The first-order wavefunction is a linear combination
of all doubly excited configurations, and their amplitudes $T_{pq}^{fg}$
are obtained by solving the equations for the MP2 residuals \citep{Saebo2002}.
The dynamic energy correction takes the form
\begin{equation}
E^{dyn}=\sum\limits _{g,f=1}^{\mathrm{N}/2}\sum\limits _{p,q>N/2}^{M}\left\langle gf\right|\left.pq\right\rangle \left[2T_{pq}^{gf}\right.\left.-T_{pq}^{fg}\right]\label{E2}
\end{equation}

where $M$ is the number of basis functions, and $\left\langle gf\right|\left.pq\right\rangle $
are the matrix elements of the two-particle interaction.

In fact, $E^{dyn}$ is the modified $E^{\left(2\right)}$ in order
to avoid double counting of the electron correlation. It is divided
into intra- and inter-pair contributions, and the amount of dynamic
correlation in each orbital $p$ is defined by functions $C_{p}$
of its occupancy, namely,
\begin{equation}
\begin{array}{c}
C_{p}^{tra}=\begin{cases}
\begin{array}{c}
\begin{array}{c}
1-4h_{p}^{2}\end{array}\\
1-4n_{p}^{2}
\end{array} & \begin{array}{c}
p\leq\mathrm{N}/2\\
p>\mathrm{N}/2
\end{array}\end{cases}\\
\:C_{p}^{ter}=\begin{cases}
\begin{array}{c}
\begin{array}{c}
1\end{array}\\
1-4h_{p}n_{p}
\end{array} & \begin{array}{c}
p\leq\mathrm{N}/2\\
p>\mathrm{N}/2
\end{array}\end{cases}
\end{array}\label{Cp}
\end{equation}
According to Eq.(\ref{Cp}), fully occupied and empty orbitals yield
a maximal contribution to dynamic correlation, whereas orbitals with
half occupancies contribute nothing. Using these functions as the
case may be (intra-pair or inter-pair), the modified off-diagonal
elements of the Fock matrix ($\tilde{\mathcal{F}}$) are defined as
\begin{equation}
\tilde{\mathcal{F}}_{pq}=\begin{cases}
C_{p}^{tra}C_{q}^{tra}\mathcal{F}_{pq}, & p,q\in\Omega_{g}\\
C_{p}^{ter}C_{q}^{ter}\mathcal{F}_{pq}, & otherwise
\end{cases}
\end{equation}
as well as modified two-electron integrals:
\begin{equation}
\widetilde{\left\langle pq\right|\left.rt\right\rangle }=\begin{cases}
C_{p}^{tra}C_{q}^{tra}C_{r}^{tra}C_{t}^{tra}\left\langle pq\right|\left.rt\right\rangle , & p,q,r,t\in\Omega_{g}\\
C_{p}^{ter}C_{q}^{ter}C_{r}^{ter}C_{t}^{ter}\left\langle pq\right|\left.rt\right\rangle , & otherwise
\end{cases}
\end{equation}
where the subspace index $g=1,...,\mathrm{N}/2$. This leads to the
following linear equation for the modified MP2 residuals:
\begin{equation}
\widetilde{\left\langle ab\right|\left.ij\right\rangle }+\left(\mathcal{F}_{aa}\right.+\mathcal{F}_{bb}-\mathcal{F}_{ii}-\left.\mathcal{F}_{jj}\right)T_{ab}^{ij}\:+\label{residual}
\end{equation}
\[
{\displaystyle \sum_{c\neq a}\mathcal{\tilde{F}}_{ac}T_{cb}^{ij}}+{\displaystyle \sum_{c\neq b}}T_{ac}^{ij}\mathcal{\tilde{F}}_{cb}-{\displaystyle \sum_{k\neq i}}\tilde{\mathcal{F}}_{ik}T_{ab}^{kj}-{\displaystyle \sum_{k\neq j}}T_{ab}^{ik}\mathcal{\tilde{F}}_{kj}=0
\]
where $i,j,k$ refer to the strong occupied NOs, and $a,b,c$ to weak
occupied ones. It should be noted that diagonal elements of the Fock
matrix ($\mathcal{F}$) are not modified. By solving this linear system
of equations the amplitudes $T_{pq}^{fg}$ are obtained, which are
inserted into the Eq. (\ref{E2}) to achieve $E^{dyn}$. 

All calculations have been carried out using the DoNOF code developed 
by M. Piris and coworkers. The procedure is simple, showing a formal 
scaling of $M^{5}$ ($M$: number of basis functions). However, our 
implementation in the molecular basis set requires also four-index 
transformation of the electron repulsion integrals, which
is a time-consuming step, though a parallel implementation of this
part of the code has substantially improved its performance. As a
result, the possibility of addressing large systems opens up.

\section{Results and Discussion}

Results are organized as follows. First, the X-H bond dissociation
energies for LiH, BH, CH$_{4}$, NH$_{3}$, H$_{2}$O and HF molecules
are studied using NOF-MP2. Next, we analyze the performance of NOF-MP2
for describing C-H bond cleavage in a variety of 20 organic molecules.
Finally, radical stabilization energies are calculated based on the
calculated C-H bond dissociation energies. In all calculations, recall
that the maximum value allowed by the basis set used is assumed for
$\mathrm{N}_{g}$ by default. In NOF-MP2(3) calculations, only three
orbitals ($\mathrm{N}_{g}=3$) above the N/2 level in each electron
pair are considered.

Geometries are taken from our previous publication \citep{Lopez2015},
which were obtained at the M06-2X level of theory \citep{Zhao2008}.
The dissociation limit is calculated by considering a frozen X-H distance
of 5$\textrm{Å}$, and optimizing the rest of internal coordinates.
At these geometries, single-point energies are evaluated at the NOF-MP2
level of theory. The correlation-consistent valence double-$\zeta$
(cc-pVDZ) or triple-$\zeta$ (cc-pVTZ) basis sets developed by Dunning
et al. \citep{Dunning1989} are used. The zero point vibrational energies
(ZPVEs) were taken from the NITS Computational Chemistry Comparison
and Benchmark Database (CCCBDB) \citep{CCCBDB}, and corresponds to
CCSD(T)/cc-pVTZ values.

We also provide the PNOF6 \citep{Piris2014c} and wavefunction-based
calculations obtained in Ref. \citep{Lopez2015}. For the latter,
an active space was defined by the distribution of two electrons in
two molecular orbitals, CASSCF(2,2) \citep{Roos1980,Siegbahn1980}.
The dynamic correlation effects were included through complete active
space second-order perturbation theory calculations, CASPT2(2,2) \citep{Anderson1992}.
MOLCAS 7.0 suite of programs \citep{Aquilante2009} was used in Ref.
\citep{Lopez2015}, for these wavefunction-based calculations.

\subsection{X-H homolytic bond cleavage }

\begin{figure}[H]
\caption{\label{fig:PNOF-DeXH}NOF-MP2 dissociation energies, in kcal/mol,
for X-H bonds ($\mathrm{X=Li,B,CH_{3},NH_{2},OH,F}$) versus experimental
ones. Calculations carried out with the cc-pVTZ basis set.\bigskip{}
}

\centering{}\includegraphics[scale=0.4]{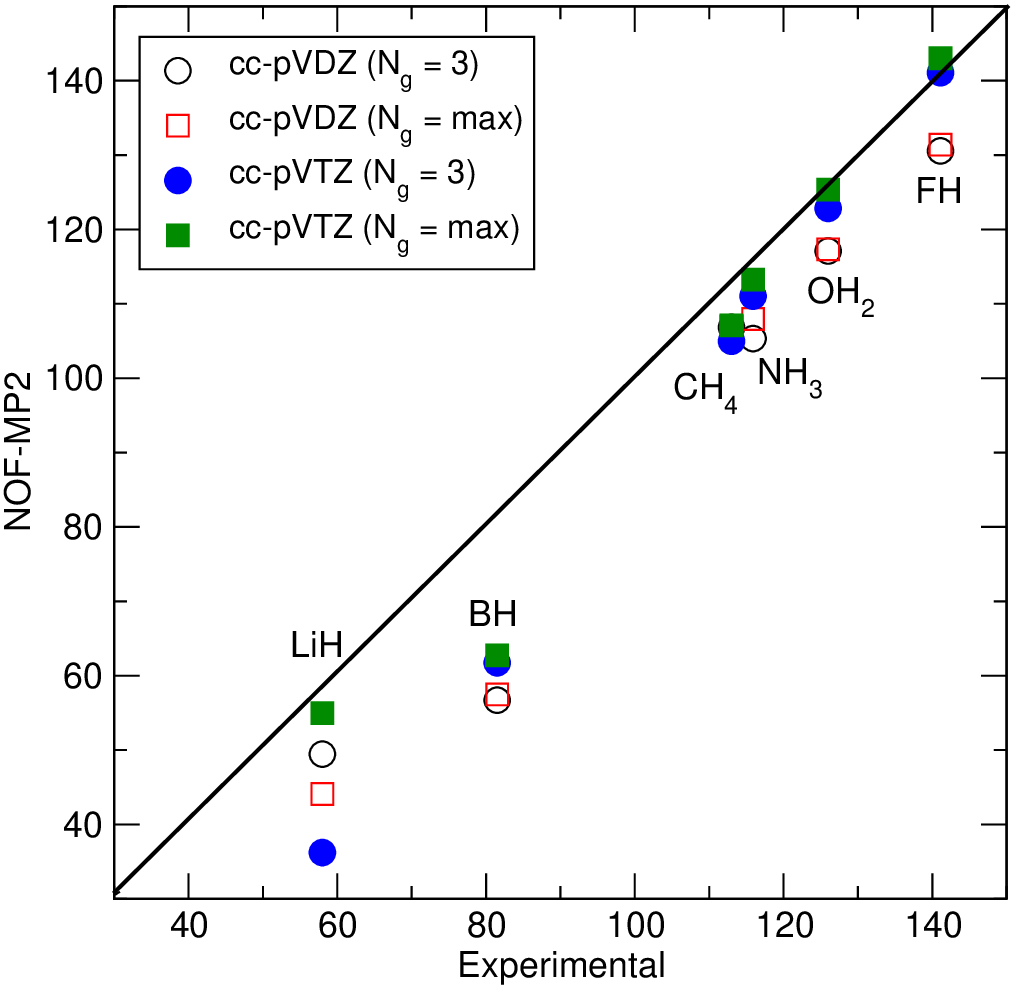}
\end{figure}

X-H bond dissociation energies were calculated according to the following
reaction:
\begin{equation}
\mathrm{XH}\rightarrow\mathrm{^{.}X}+\mathrm{H}^{.}
\end{equation}
with X = Li, B, CH$_{3}$, NH$_{2}$, OH, F. The results are presented
in Figure \ref{fig:PNOF-DeXH} and Table \ref{tab:PNOF-DeXH}. The
different hydrides considered expand a wide range of dissociation
energies, from 58.0 kcal/mol for LiH to 141.1 kcal/mol for FH. The
ordering in dissociation energies is $\mathrm{LiH}<\mathrm{BH}<\mathrm{CH_{4}}<\mathrm{NH_{3}}<\mathrm{H_{2}O}<\mathrm{FH}$.
In general, NOF-MP2 reproduces satisfactorily these trends.

Let us focus our attention, for example, on the delicate case of the
CH$_{4}$/NH$_{3}$ ordering. The difference in experimental dissociation
energies for these two molecules is very small, only 2.9 kcal/mol
with $\mathrm{NH_{3}}$ having a higher dissociation energy. NOF-MP2
is able to reproduce the correct ordering $\mathrm{CH_{4}<NH_{3}}$,
except for \textcolor{black}{NOF-MP2(3)/cc-pVDZ}. It should be noted
that CASSCF(2,2) and PNOF6 gives the reverse order, whereas CASPT2(2,2)
recovers the right trend.

\begin{table*}
\caption{\label{tab:PNOF-DeXH} Dissociation energies, in kcal/mol, calculated
from single-point energies$^{a}$. ZPVEs were added$^{b}$ to the
experimental dissociation energies \citep{Ervin2002}. PNOF6, CASSCF(2,2)
and CASPT2(2,2) results are taken from Ref. \citep{Lopez2015}.\bigskip{}
}

\begin{tabular}{cccccccc}
\hline 
 & \multicolumn{1}{c|}{$\mathrm{LiH\rightarrow{}^{.}Li+H^{.}}$} & \multicolumn{1}{c|}{$\mathrm{BH\rightarrow{}^{.}B+H^{.}}$} & \multicolumn{1}{c|}{$\mathrm{CH_{4}\rightarrow^{.}CH_{3}+H^{.}}$} & \multicolumn{1}{c|}{$\mathrm{NH_{3}\rightarrow^{.}NH_{2}+H^{.}}$} & \multicolumn{1}{c|}{$\mathrm{H_{2}O\rightarrow{}^{.}OH+H^{.}}$} & \multicolumn{1}{c}{$\mathrm{FH\rightarrow{}^{.}F+H^{.}}$} & MAE\tabularnewline
\hline 
 & \multicolumn{7}{c}{cc-pVDZ}\tabularnewline
\hline 
 &  &  &  &  &  &  & \tabularnewline
PNOF6$^{c}$ & 42.8 & 78.4 & 104.6 & 102.4 & 106.7 & 113.2 & 14.6\tabularnewline
PNOF6(3)$^{d}$ & 48.5 & 89.4 & 112.9 & 111.2 & 114.7 & 119.8 & 9.1\tabularnewline
\textcolor{black}{NOF-MP2}$^{c}$ & \textcolor{black}{44.1} & \textcolor{black}{57.5} & \textcolor{black}{106.9} & \textcolor{black}{108.0} & \textcolor{black}{117.3} & \textcolor{black}{131.4} & 11.7\tabularnewline
\textcolor{black}{NOF-MP2(3)}$^{d}$ & \textcolor{black}{49.5} & \textcolor{black}{56.7} & \textcolor{black}{106.8} & \textcolor{black}{105.3} & \textcolor{black}{117.1} & \textcolor{black}{130.6} & 11.6\tabularnewline
CASSCF(2,2) & 42.8 & 78.2 & 97.3 & 95.1 & 99.4 & 107.8 & 19.2\tabularnewline
CASPT2(2,2) & 49.2 & 78.7 & 106.6 & 106.8 & 114.8 & 126.6 & 8.8\tabularnewline
\hline 
 & \multicolumn{7}{c}{cc-pVTZ}\tabularnewline
\hline 
 &  &  &  &  &  &  & \tabularnewline
PNOF6$^{c}$ & 44.1 & 80.7 & 105.2 & 104.8 & 110.4 & 119.5 & 11.8\tabularnewline
PNOF6(3)$^{d}$ & 51.3 & 92.6 & 113.3 & 114.4 & 120.1 & 127.8 & 6.5\tabularnewline
\textcolor{black}{NOF-MP2}$^{c}$ & \textcolor{black}{55.0} & \textcolor{black}{62.7} & \textcolor{black}{107.1} & \textcolor{black}{113.3} & \textcolor{black}{125.4} & \textcolor{black}{143.0} & 5.5\tabularnewline
\textcolor{black}{NOF-MP2(3)}$^{d}$ & \textcolor{black}{36.2} & \textcolor{black}{61.7} & \textcolor{black}{105.0} & \textcolor{black}{111.0} & \textcolor{black}{122.8} & \textcolor{black}{141.1} & 9.6\tabularnewline
CASSCF(2,2) & 44.0 & 81.1 & 98.0 & 97.3 & 102.7 & 113.4 & 16.5\tabularnewline
CASPT2(2,2) & 53.4 & 81.7 & 109.6 & 112.2 & 122.0 & 136.9 & 3.4\tabularnewline
\hline 
Exp. & 58.0 & 81.5 & 113.0 & 115.9 & 126.0 & 141.1 & \tabularnewline
\hline 
\end{tabular}\bigskip{}

\begin{raggedright}
{\footnotesize{}{a) Calculation of the energy at the dissociation
limit were done at a X-H distance of 5 $\textrm{Å}$.}}{\footnotesize\par}
\par\end{raggedright}
\begin{raggedright}
{\footnotesize{}{b) ZPVEs were taken from NITS CCCBDB \citep{CCCBDB},
and corresponds to CCSD(T)/cc-pVTZ values.}}{\footnotesize\par}
\par\end{raggedright}
\begin{raggedright}
{\footnotesize{}{c) In these calculations, N$_{g}$ is the maximum
allowed value determined by the basis set used.}}{\footnotesize\par}
\par\end{raggedright}
\raggedright{}{\footnotesize{}{d) In these calculations, N$_{g}$
= 3.}}{\footnotesize\par}
\end{table*}

It is well known that to reach the experimental values \LyXZeroWidthSpace \LyXZeroWidthSpace we
must go to the complete basis set limit. Therefore, taking into account
the moderate basis sets used here, we can say that a good semi-quantitative
agreement has been achieved with the experimental data by the NOF-MP2
method. In general, NOF-MP2 shows an intermediate performance between
the CASSCF(2,2) and CASPT2(2,2) methods, and a significant improvement
with respect to the previously tested PNOF6.

For the six reactions considered, a mean absolute error (MAE) of 5.5
kcal/mol is obtained at NOF-MP2/cc-pVTZ level of theory. NOF-MP2(3)/cc-pVTZ,
leads to a higher MAE, namely 9.6 kcal/mol, but this is mainly due
to LiH case. For the latter, only one effective pair appears so more
$\mathrm{N}_{g}$ orbitals are needed in order to describe properly
the dominant intra-pair electron correlation \citep{Piris2013b} in
this system. On the other hand, for $\mathrm{CH_{4},NH_{3},H_{2}O}$
and FH, NOF-MP2(3) yield very reasonable results. Therefore, we can
say that $\mathrm{N}_{g}=3$ is a good compromise for the characterization
of the electron pairs, except for small systems like LiH and BH.

Comparing the performance of NOF-MP2 with wavefunction methods, it
is clear that NOF-MP2 and NOF-MP2(3) show a better performance than
CASSCF(2,2) (MAE=16.5 kcal/mol with the cc-pVTZ basis set). Introduction
of dynamical electron correlation at the CASPT2(2,2) level of theory,
reduces the MAE to 3.4 kcal/mol, however, if we reduce the set to
CH$_{4}$, NH$_{3}$, H$_{2}$O and FH molecules, there is a similar
performance of NOF-MP2 with respect to CASPT2(2,2) method.

\subsection{Hydrogen Abstraction in a Dataset of 20 Organic Molecules}

\subsubsection{Dissociation Energies}

As in our previous work \citep{Lopez2015}, we have considered a dataset
of 20 organic molecules to evaluate the performance of NOF-MP2 for
the C-H bond dissociation energy ($\mathrm{D_{e}^{CH}}$). The selected
set covers a wide range of $\mathrm{D_{e}^{CH}}$ values, from 95.6
kcal/mol (H$_{2}$CO) to 141.8 kcal/mol (C$_{2}$H$_{2}$), showing
the sensitivity of the C-H bond to different chemical environments.
We have considered functional groups with different degree of electron
withdrawing/donating ability (-F, -OH, -NO$_{2}$, -CN, -CH$_{3}$,
...), aromaticity (-C$_{6}$H$_{5}$), variety of C-X bonds (HCN,
H$_{2}$CO, CH$_{3}$NO$_{2}$, CH$_{3}$CF$_{3}$, ...), different
chain lengths (CH$_{4}$, CH$_{3}$CH$_{3}$, CH$_{3}$CH$_{2}$CH$_{3}$)
and different C-C bond orders, single (as in CH$_{3}$CH$_{3}$),
double (as in C$_{2}$H$_{4}$) and triple (as in C$_{2}$H$_{2}$).
We have decided to use the cc-pVDZ basis set due to the large number
of compounds to be treated.

\begin{table*}
\caption{\label{tab:CH-De}C-H Bond Dissociation energies, in kcal/mol, for
a dataset of 20 organic molecules. ZPVEs at the M062X/cc-pVTZ level
of theory were added to the experimental dissociation energies \citep{Henry2001,Blanksby2003}.
In case of CH$_{4}$, this leads to a experimental D$_{e}$ of 112.7
kcal/mol, 0.3 kcal/mol lower than the value estimated in Table \ref{tab:PNOF-DeXH}.
PNOF6(3), CASSCF(2,2) and CASPT2(2,2) data is taken from Ref. \citep{Lopez2015}.
Calculations carried out with the cc-pVDZ basis set, and considering
the X-H distance of 5 $\textrm{Å}$ as the dissociation limit. \bigskip{}
}

\textcolor{black}{\small{}}%
\begin{tabular}{l|cccccc}
 &  & \textcolor{black}{\small{}PNOF6(3) } & \textcolor{black}{\small{}CASSCF(2,2) } & \textcolor{black}{\small{}CASPT2(2,2)} & \textcolor{black}{\small{}NOF-MP2(3)} & \textcolor{black}{\small{}Exp }\tabularnewline
\hline 
\textcolor{black}{\small{}$\mathrm{CH_{4}}$} &  & \textcolor{black}{\small{}112.8} & \textcolor{black}{\small{}97.3 } & \textcolor{black}{\small{}106.6} & \textcolor{black}{\small{}104.5} & \textcolor{black}{\small{}112.7 }\tabularnewline
\textcolor{black}{\small{}$\mathrm{CH_{3}CH_{3}}$} &  & \textcolor{black}{\small{}111.2 } & \textcolor{black}{\small{}96.0 } & \textcolor{black}{\small{}104.2} & \textcolor{black}{\small{}103.4} & \textcolor{black}{\small{}109.7 }\tabularnewline
\textcolor{black}{\small{}$\mathrm{CH_{3}CH_{2}CH_{3}}$$^{a}$} &  & \textcolor{black}{\small{}110.8 } & \textcolor{black}{\small{}94.5 } & \textcolor{black}{\small{}102.5} & \textcolor{black}{\small{}103.9} & \textcolor{black}{\small{}106.9 }\tabularnewline
\textcolor{black}{\small{}$\mathrm{CH_{3}CH_{2}CH_{3}}$$^{b}$} &  & \textcolor{black}{\small{}112.7 } & \textcolor{black}{\small{}96.5 } & \textcolor{black}{\small{}104.8} & \textcolor{black}{\small{}105.8} & \textcolor{black}{\small{}108.5 }\tabularnewline
\textcolor{black}{\small{}$\mathrm{CH_{3}F}$} &  & \textcolor{black}{\small{}111.4 } & \textcolor{black}{\small{}97.9 } & \textcolor{black}{\small{}103.0} & \textcolor{black}{\small{}108.5} & \textcolor{black}{\small{}108.7 }\tabularnewline
\textcolor{black}{\small{}$\mathrm{CF_{2}H_{2}}$ } &  & \textcolor{black}{\small{}118.3 } & \textcolor{black}{\small{}100.3 } & \textcolor{black}{\small{}105.4} & \textcolor{black}{\small{}116.1} & \textcolor{black}{\small{}111.8 }\tabularnewline
\textcolor{black}{\small{}$\mathrm{CF_{3}H}$ } &  & \textcolor{black}{\small{}121.2 } & \textcolor{black}{\small{}101.7 } & \textcolor{black}{\small{}107.3} & \textcolor{black}{\small{}117.6} & \textcolor{black}{\small{}113.5 }\tabularnewline
\textcolor{black}{\small{}$\mathrm{CH_{3}OH}$ } &  & \textcolor{black}{\small{}113.4 } & \textcolor{black}{\small{}94.5 } & \textcolor{black}{\small{}99.1} & \textcolor{black}{\small{}106.7} & \textcolor{black}{\small{}103.2 }\tabularnewline
\textcolor{black}{\small{}$\mathrm{CH_{3}COH}$ } &  & \textcolor{black}{\small{}110.6 } & \textcolor{black}{\small{}91.7 } & \textcolor{black}{\small{}97.4} & \textcolor{black}{\small{}101.3} & \textcolor{black}{\small{}100.8 }\tabularnewline
\textcolor{black}{\small{}$\mathrm{H_{2}CO}$ } &  & \textcolor{black}{\small{}112.0 } & \textcolor{black}{\small{}81.6 } & \textcolor{black}{\small{}87.2} & \textcolor{black}{\small{}101.6} & \textcolor{black}{\small{}95.6 }\tabularnewline
\textcolor{black}{\small{}$\mathrm{CH_{3}OCH_{3}}$ } &  & \textcolor{black}{\small{}113.7 } & \textcolor{black}{\small{}95.6 } & \textcolor{black}{\small{}100.7} & \textcolor{black}{\small{}108.0} & \textcolor{black}{\small{}102.5 }\tabularnewline
\textcolor{black}{\small{}$\mathrm{CH_{3}COOH}$ } &  & \textcolor{black}{\small{}113.8 } & \textcolor{black}{\small{}92.9 } & \textcolor{black}{\small{}100.6} & \textcolor{black}{\small{}102.5} & \textcolor{black}{\small{}100.8 }\tabularnewline
\textcolor{black}{\small{}$\mathrm{C_{2}H_{4}}$ } &  & \textcolor{black}{\small{}128.9 } & \textcolor{black}{\small{}103.7 } & \textcolor{black}{\small{}115.8} & \textcolor{black}{\small{}116.9} & \textcolor{black}{\small{}119.3 }\tabularnewline
\textcolor{black}{\small{}$\mathrm{C_{2}H_{2}}$ } &  & \textcolor{black}{\small{}147.9 } & \textcolor{black}{\small{}127.0 } & \textcolor{black}{\small{}149.7} & \textcolor{black}{\small{}143.4} & \textcolor{black}{\small{}141.8 }\tabularnewline
\textcolor{black}{\small{}$\mathrm{CH_{3}CCH}$ } &  & \textcolor{black}{\small{}112.1 } & \textcolor{black}{\small{}92.1 } & \textcolor{black}{\small{}94.5} & \textcolor{black}{\small{}101.7} & \textcolor{black}{\small{}97.1 }\tabularnewline
\textcolor{black}{\small{}$\mathrm{HCN}$ } &  & \textcolor{black}{\small{}141.2 } & \textcolor{black}{\small{}126.5 } & \textcolor{black}{\small{}125.8} & \textcolor{black}{\small{}127.2} & \textcolor{black}{\small{}132.5 }\tabularnewline
\textcolor{black}{\small{}$\mathrm{CH_{3}CN}$ } &  & \textcolor{black}{\small{}119.4 } & \textcolor{black}{\small{}94.8 } & \textcolor{black}{\small{}98.3} & \textcolor{black}{\small{}104.2} & \textcolor{black}{\small{}103.9 }\tabularnewline
\textcolor{black}{\small{}$\mathrm{CH_{3}NO_{2}}$ } &  & \textcolor{black}{\small{}127.5 } & \textcolor{black}{\small{}98.2 } & \textcolor{black}{\small{}103.3} & \textcolor{black}{\small{}108.5} & \textcolor{black}{\small{}106.5 }\tabularnewline
\textcolor{black}{\small{}$\mathrm{CF_{3}CH_{3}}$ } &  & \textcolor{black}{\small{}114.6 } & \textcolor{black}{\small{}99.1 } & \textcolor{black}{\small{}107.9} & \textcolor{black}{\small{}108.7} & \textcolor{black}{\small{}113.8 }\tabularnewline
\textcolor{black}{\small{}$\mathrm{C_{6}H_{6}}$ } &  & \textcolor{black}{\small{}129.8 } & \textcolor{black}{\small{}101.6 } & \textcolor{black}{\small{}114.6} & \textcolor{black}{\small{}124.3} & \textcolor{black}{\small{}120.5 }\tabularnewline
\textcolor{black}{\small{}$\mathrm{C_{6}H_{6}CH_{3}}$ } &  & \textcolor{black}{\small{}111.9 } & \textcolor{black}{\small{}89.6 } & \textcolor{black}{\small{}103.7} & \textcolor{black}{\small{}103.4} & \textcolor{black}{\small{}96.1 }\tabularnewline
\hline 
\textcolor{black}{\small{}MAE} &  & \textcolor{black}{\small{}9.0 } & \textcolor{black}{\small{}11.1 } & \textcolor{black}{\small{}5.0} & \textcolor{black}{\small{}3.7} & \tabularnewline
\hline 
\end{tabular}\bigskip{}

\begin{centering}
{\footnotesize{}{a) hydrogen abstraction from the terminal -CH$_{3}$
group}}{\footnotesize\par}
\par\end{centering}
\centering{}{\footnotesize{}{b) hydrogen abstraction from the central
-CH$_{2}$- group }}{\footnotesize\par}
\end{table*}

\begin{figure}
\caption{\label{fig:CH-De}C-H bond dissociation energies, in kcal/mol, for
the Table \ref{tab:CH-De} dataset of 20 organic molecules. All calculations
were done with the cc-pVDZ basis set. The dissociation limit distance
was taken as 5$\textrm{Å}$.\bigskip{}
}

\centering{}\includegraphics[scale=0.4]{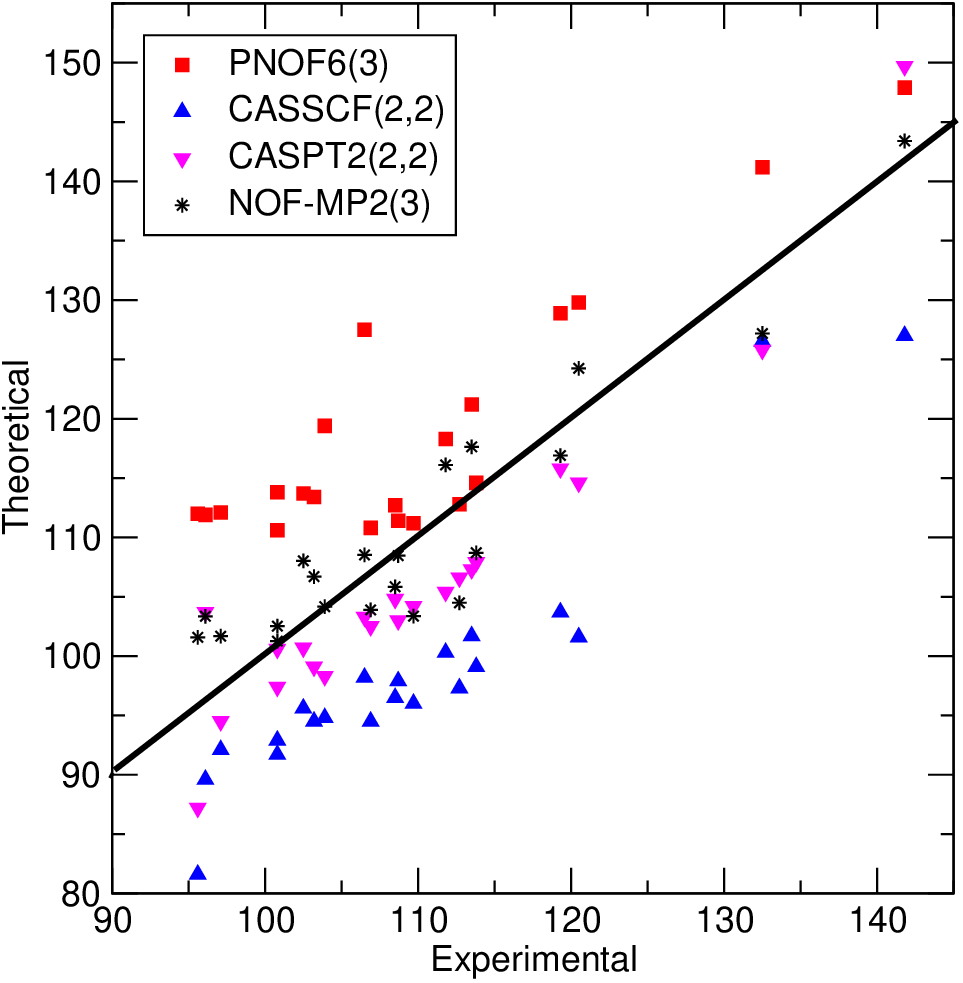}
\end{figure}

\begin{figure}
{\footnotesize{}\caption{\label{fig:RSE}Radical Stabilization Energies, in kcal/mol, based
on the combination of dissociation energies of Table \ref{tab:CH-De}.
All calculations were done with the cc-pVDZ basis set. The mean absolute
errors with respect to the experimental values are 5.5 kcal/mol, 4.4
kcal/mol, and 4.1 kcal/mol for NOF-MP2(3), CASSCF(2,2), and CASPT2(2,2),
respectively.\bigskip{}
}
}{\footnotesize\par}
\begin{centering}
{\footnotesize{}\includegraphics[scale=0.4]{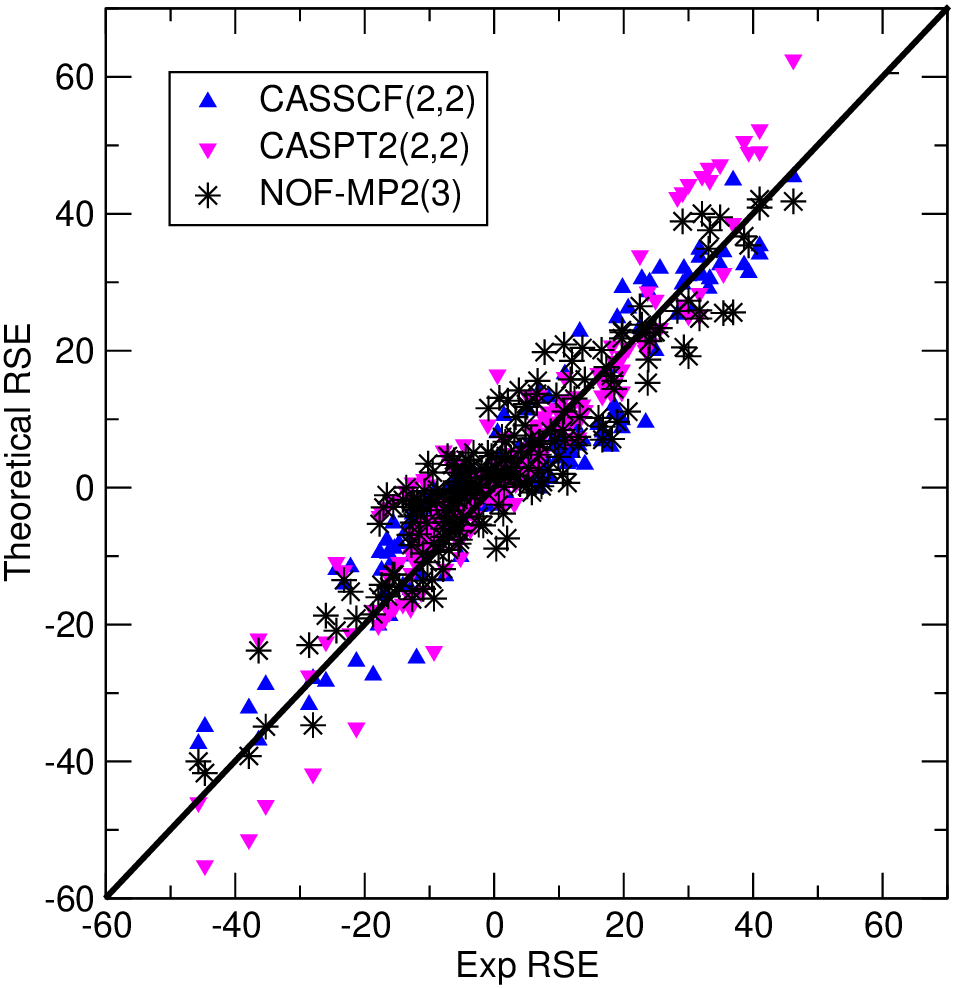}}{\footnotesize\par}
\par\end{centering}
\centering{}%
\begin{tabular}{c|c|c}
 & {\footnotesize{}{linear fit}} & {\footnotesize{}{r}}\tabularnewline
\hline 
{\footnotesize{}{CASSCF(2,2)}} & {\footnotesize{}{$y=1.0426+0.8561*x$ }} & {\footnotesize{}{0.9482}}\tabularnewline
{\footnotesize{}{CASPT2(2,2)}} & {\footnotesize{}{$y=1.6895+1.0358*x$ }} & {\footnotesize{}{0.9482}}\tabularnewline
\textcolor{black}{\footnotesize{}{NOF-MP2(3)}} & \textcolor{black}{\footnotesize{}{$y=2.1468+0.8408*x$ }} & \textcolor{black}{\footnotesize{}{0.9314}}\tabularnewline
\end{tabular}
\end{figure}

The results can be found in Table \ref{tab:CH-De} and Figure \ref{fig:CH-De}.
The agreement between NOF-MP2(3) and experimental values is remarkable,
with a MAE of 3.7 kcal/mol, even smaller than the MAE for the very
accurate CASPT2(2,2) method, namely 5.0 kcal/mol. Notice that previously
tested PNOF6(3) method has a MAE of 9.0 kcal/mol, slightly better
than CASSCF(2,2), 11.1 kcal/mol. Thus, NOF-MP2(3) method allows for
a quantitative description of these dissociation energies, with a
similar degree of accuracy as CASPT2(2,2).

Specifically, NOF-MP2(3) is able to reproduce important trends in
C-H bond energies. For instance, the experimental $\mathrm{D_{e}^{CH}}$
increases in the following order \citep{Blanksby2003}: CH$_{3}$CH$_{3}$
(109.7) < C$_{2}$H$_{4}$ (119.3) < C$_{2}$H$_{2}$ (141.8) . NOF-MP2(3)
is able to reproduce properly this trend, namely, CH$_{3}$CH$_{3}$
(103.4) < C$_{2}$H$_{4}$ (116.9) < C$_{2}$H$_{2}$ (143.4).

The effect of aromaticity can be inferred from the comparison of these
dissociation energies with that of the phenyl C-H bond. C$_{6}$H$_{6}$,
with a formal 1.5 C-C bond order, shows a high dissociation energy
(120.5 kcal/mol) even slightly larger than that observed (119.3 kcal/mol)
in C$_{2}$H$_{4}$, with a formal bond order of 2. This is a clear
signature of aromaticity in C$_{6}$H$_{6}$, partially lost upon
hydrogen abstraction and radical formation. NOF-MP2(3) yields larger
values of bond dissociation energies for benzene than for ethene,
with values of 124.3 kcal/mol and 116.9 kcal/mol, respectively\textcolor{black}{\small{}.}{\small\par}

In the case of the benzylic C-H bond (C$_{6}$H$_{6}$CH$_{2}$-H),
the effect of the aromaticity works in the opposite direction. In
this case, the C-H cleavage does not break the aromaticity, furthermore,
the radical itself is stabilized by the aromatic character of the
phenyl ring, and consequently, one obtains a much lower $\mathrm{D_{e}^{CH}}$
than for C$_{6}$H$_{6}$, namely 96.1 kcal/mol versus 120.5 kcal/mol.
NOF-MP2(3) correctly describes this effect, $\mathrm{D_{e}^{CH}}$
for the benzylic C-H bond (103.4 kcal/mol) is also much lower than
for the phenyl C-H bond (124.3 kcal/mol) at a magnitude very similar
to the experimental value. It is remarkable the right description
of aromatic radical stabilization by the NOF-MP2 method, since aromatic
stabilization is key to describe radical stability in chemistry.

The chain length is also a factor influencing the C-H bond strength
\citep{Vereecken2001,Carstensen2007,Huynh2010}. It is known that
a larger chain stabilizes the resulting radical: observe the first
3 lines of Table \ref{tab:CH-De}. However, NOF-MP2(3) exhibits a
poorer sensitivity of radical stability towards chain-lengths with
a similar $\mathrm{D_{e}^{CH}}$ for these three molecules. On the
other hand, if we consider the same alkane, CH$_{3}$CH$_{2}$CH$_{3}$,
and measure both possibilities for hydrogen abstraction, namely, from
the central -CH$_{2}$- or from the terminal -CH$_{3}$ group, NOF-MP2(3)
correctly reproduces the more favorable hydrogen abstraction from
the central carbon by 1.9 kcal/mol.

There is also a sizable effect in hydrogen abstraction upon the inclusion
of electron withdrawing groups. For instance, fluorination \citep{Korchowiec2002}
and oxidation \citep{Blanksby2003} of methane tend to alter the dissociation
energy of the C-H bond. Regarding fluorination, a decrease of $\mathrm{D_{e}^{CH}}$
is observed upon the inclusion of a first flour, from 112.7 kcal/mol
(CH$_{4}$) to 108.7 kcal/mol in (CH$_{3}$F). However, upon higher
degree of fluorination in the fluoromethane, $\mathrm{D_{e}^{CH}}$
increases again, 111.8 kcal/mol in CF$_{2}$H$_{2}$ and 113.5 kcal/mol
in CF$_{3}$H. NOF-MP2(3) yields a higher $\mathrm{D_{e}^{CH}}$ for
CH$_{3}$F (108.5 kcal/mol) than for CH$_{4}$ (104.5 kcal/mol). Nevertheless,
NOF-MP2(3) describes the proper trend in increasing $\mathrm{D_{e}^{CH}}$
with the degree of fluorination in fluoromethane, namely, CH3F (108.5
kcal/mol) < CH$_{2}$F$_{2}$ (116.1 kcal/mol) < CHF$_{3}$ (117.6
kcal/mol).

With respect to the oxidation of a methyl group, NOF-MP2(3) gives
the right trend. For instance, in going from CH$_{3}$OH to H$_{2}$CO,
there is an important reduction in C-H bond strength, from 103.2 kcal/mol
to 95.6 kcal/mol. NOF-MP2(3) yields a similar, although more discrete
reduction, of $\mathrm{D_{e}^{CH}}$ from 106.7 kcal/mol to 101.6
kcal/mol.

In general, we can conclude that NOF-MP2(3) represents an accurate
balance between dynamical and non-dynamical electron correlation for
this set of molecules, yielding $\mathrm{D_{e}^{CH}}$ values that
are of the CASPT2(2,2) quality.

\subsubsection{Radical Stabilization Energies}

RSEs are defined as the energy change in the isodesmic reaction for
hydrogen abstraction \citep{Wood2005,Menon2007,Menon2008} of Eq.
(\ref{eq:isodesmicEq}). Thus, the RSE for a pair X,Y is defined as

\begin{equation}
\mathrm{RSE^{XY}}=\mathrm{D_{e}^{XH}}-\mathrm{D_{e}^{YH}}
\end{equation}

For the dataset of 21 dissociation energies of Table \ref{tab:CH-De},
there are 210 possible combinations of RSEs. It provides with an extensive
dataset for the determination of the suitability of a given method
to estimate the effect of the substituents on the radical stability
in organic molecules. The results for NOF-MP2(3) are summarized in
Fig. \ref{fig:RSE}, compared to the performance of wavefunction methods
such as CASSCF(2,2) and CASPT2(2,2). In general, there is a reasonable
agreement with experimental RSEs for NOF-MP2(3) with a MAE of 5.5
kcal/mol. Slightly better values are obtained for CASSCF(2,2) (4.4
kcal/mol) and CASPT2(2,2) (4.1 kcal/mol) levels of theory.

Another way to compare the results with respect to experimental values
is to calculate the linear fit of the theoretical versus the experimental
values, and determine the correlation coefficient ($r$). In this
sense, NOF-MP2(3) shows a similar performance to the CASSCF(2,2) and
CAS2PT2(2,2) methods with an $r$ of 0.9314 versus a value of 0.9482
for both wavefunction methods. In summary, taking into account the
large number of hydrogen abstraction reactions considered, the correlation
between NOF-MP2(3) and experimental data is highly satisfactory, yielding
a quantitative agreement with respect to well established wavefunction
methods such as CASPT2(2,2), and providing results close to chemical
accuracy.

\section{Conclusions}

The recently proposed \textcolor{black}{parameter-free natural orbital
functional second-order Møller\textendash Plesset (NOF-MP2) method
has been }applied to the description of radical formation reactions,
a delicate problem in quantum chemistry. The application of NOF-MP2(3)
to the calculation of the C-H bond dissociation energy in a dataset
of 20 organic molecules, and the estimation of the corresponding radical
stabilization energies support the use of NOF-MP2(3) as a quantitative
theory for the description of these important set of reactions. Comparison
of NOF-MP2 with experimental data reveals a similar performance of
NOF-MP2 to well-established wavefunction methods such as CASPT2 for
these type of problems. We conclude that NOF-MP2 is capable of recovering
both dynamical and non-dynamical electron correlation effects in this
type of systems. NOF-MP2 is a global electron correlation method for
the description of radical stability, which provides results close
to chemical accuracy as the widely used and well-established CASPT2
wavefunction method.

\begin{acknowledgments}
Financial support comes from Ministerio de Economía y Competitividad
(Ref. CTQ2015-67608-P). The authors thank for technical and human
support provided by IZO-SGI SGIker of UPV/EHU and European funding
(ERDF and ESF).
\end{acknowledgments}


\begin{thebibliography}{48}
\expandafter\ifx\csname natexlab\endcsname\relax\def\natexlab#1{#1}\fi
\expandafter\ifx\csname bibnamefont\endcsname\relax
  \def\bibnamefont#1{#1}\fi
\expandafter\ifx\csname bibfnamefont\endcsname\relax
  \def\bibfnamefont#1{#1}\fi
\expandafter\ifx\csname citenamefont\endcsname\relax
  \def\citenamefont#1{#1}\fi
\expandafter\ifx\csname url\endcsname\relax
  \def\url#1{\texttt{#1}}\fi
\expandafter\ifx\csname urlprefix\endcsname\relax\def\urlprefix{URL }\fi
\providecommand{\bibinfo}[2]{#2}
\providecommand{\eprint}[2][]{\url{#2}}

\bibitem[{\citenamefont{Piris}(2007)}]{Piris2007}
\bibinfo{author}{\bibfnamefont{M.}~\bibnamefont{Piris}}, in
  \emph{\bibinfo{booktitle}{Reduced-Density-Matrix Mechanics: with applications
  to many-electron atoms and molecules}}, edited by
  \bibinfo{editor}{\bibfnamefont{D.~A.} \bibnamefont{Mazziotti}}
  (\bibinfo{publisher}{John Wiley and Sons}, \bibinfo{address}{Hoboken, New
  Jersey, USA}, \bibinfo{year}{2007}), chap.~\bibinfo{chapter}{14}, pp.
  \bibinfo{pages}{387--427}.

\bibitem[{\citenamefont{Piris and Ugalde}(2014)}]{Piris2014a}
\bibinfo{author}{\bibfnamefont{M.}~\bibnamefont{Piris}} \bibnamefont{and}
  \bibinfo{author}{\bibfnamefont{J.~M.} \bibnamefont{Ugalde}},
  \bibinfo{journal}{Int. J. Quantum Chem.} \textbf{\bibinfo{volume}{114}},
  \bibinfo{pages}{1169} (\bibinfo{year}{2014}).

\bibitem[{\citenamefont{Pernal and Giesbertz}(2016)}]{Pernal2016}
\bibinfo{author}{\bibfnamefont{K.}~\bibnamefont{Pernal}} \bibnamefont{and}
  \bibinfo{author}{\bibfnamefont{K.~J.~H.} \bibnamefont{Giesbertz}},
  \bibinfo{journal}{Top Curr Chem} \textbf{\bibinfo{volume}{368}},
  \bibinfo{pages}{125} (\bibinfo{year}{2016}).

\bibitem[{\citenamefont{Piris}(2017)}]{Piris2017}
\bibinfo{author}{\bibfnamefont{M.}~\bibnamefont{Piris}},
  \bibinfo{journal}{Phys. Rev. Lett.} \textbf{\bibinfo{volume}{119}},
  \bibinfo{pages}{063002} (\bibinfo{year}{2017}).

\bibitem[{\citenamefont{Gilbert}(1975)}]{Gilbert1975}
\bibinfo{author}{\bibfnamefont{T.~L.} \bibnamefont{Gilbert}},
  \bibinfo{journal}{Phys. Rev. B} \textbf{\bibinfo{volume}{12}},
  \bibinfo{pages}{2111} (\bibinfo{year}{1975}).

\bibitem[{\citenamefont{Levy}(1979)}]{Levy1979}
\bibinfo{author}{\bibfnamefont{M.}~\bibnamefont{Levy}}, \bibinfo{journal}{Proc.
  Natl. Acad. Sci. USA} \textbf{\bibinfo{volume}{76}}, \bibinfo{pages}{6062}
  (\bibinfo{year}{1979}).

\bibitem[{\citenamefont{Valone}(1980)}]{Valone1980}
\bibinfo{author}{\bibfnamefont{S.~M.} \bibnamefont{Valone}},
  \bibinfo{journal}{J. Chem. Phys.} \textbf{\bibinfo{volume}{73}},
  \bibinfo{pages}{1344} (\bibinfo{year}{1980}).

\bibitem[{\citenamefont{Coleman}(1963)}]{Coleman1963}
\bibinfo{author}{\bibfnamefont{A.~J.} \bibnamefont{Coleman}},
  \bibinfo{journal}{Rev. Mod. Phys.} \textbf{\bibinfo{volume}{35}},
  \bibinfo{pages}{668} (\bibinfo{year}{1963}).

\bibitem[{\citenamefont{Piris}(2018{\natexlab{a}})}]{Piris2018}
\bibinfo{author}{\bibfnamefont{M.}~\bibnamefont{Piris}}, in
  \emph{\bibinfo{booktitle}{Many-body approaches at different scales: a tribute
  to N. H. March on the occasion of his 90th birthday}}, edited by
  \bibinfo{editor}{\bibfnamefont{G.~G.~N.} \bibnamefont{Angilella}}
  \bibnamefont{and} \bibinfo{editor}{\bibfnamefont{C.}~\bibnamefont{Amovilli}}
  (\bibinfo{publisher}{Springer}, \bibinfo{address}{New York},
  \bibinfo{year}{2018}{\natexlab{a}}), chap.~\bibinfo{chapter}{22}, pp.
  \bibinfo{pages}{283--300}.

\bibitem[{\citenamefont{{Piris, M}}(2013)}]{Piris2013b}
\bibinfo{author}{\bibnamefont{{Piris, M}}}, \bibinfo{journal}{Int. J. Quantum
  Chem.} \textbf{\bibinfo{volume}{113}}, \bibinfo{pages}{620}
  (\bibinfo{year}{2013}).

\bibitem[{\citenamefont{{M. Piris}}(2018)}]{Piris2018b}
\bibinfo{author}{\bibnamefont{{M. Piris}}}, \bibinfo{journal}{Phys. Rev. A}
  \textbf{\bibinfo{volume}{98}}, \bibinfo{pages}{022504}
  (\bibinfo{year}{2018}).

\bibitem[{\citenamefont{Saebo}(2002)}]{Saebo2002}
\bibinfo{author}{\bibfnamefont{S.}~\bibnamefont{Saebo}}, in
  \emph{\bibinfo{booktitle}{Computational Chemistry: Reviews of Current Trends,
  Vol. 7}} (\bibinfo{year}{2002}), pp. \bibinfo{pages}{63--87}.

\bibitem[{\citenamefont{{Valko, M.; Rhodes C. J.; Moncol J.; Izakovic, M.;
  Mazur, M.}}(2006)}]{Valko2006}
\bibinfo{author}{\bibnamefont{{Valko, M.; Rhodes C. J.; Moncol J.; Izakovic,
  M.; Mazur, M.}}}, \bibinfo{journal}{Chem-Biological Int.}
  \textbf{\bibinfo{volume}{160}}, \bibinfo{pages}{1} (\bibinfo{year}{2006}).

\bibitem[{\citenamefont{{Valko, M.; Leibfritz, D.; Moncol, J.; Cronin, M.T.D.;
  Mazur, M.; Telser, J.}}(2007)}]{Valko2007}
\bibinfo{author}{\bibnamefont{{Valko, M.; Leibfritz, D.; Moncol, J.; Cronin,
  M.T.D.; Mazur, M.; Telser, J.}}}, \bibinfo{journal}{Int. J. Biochem. and Cell
  Bio.} \textbf{\bibinfo{volume}{39}}, \bibinfo{pages}{44}
  (\bibinfo{year}{2007}).

\bibitem[{\citenamefont{{Breher, F.}}(2007)}]{Breher2007}
\bibinfo{author}{\bibnamefont{{Breher, F.}}}, \bibinfo{journal}{Coord. Chem.
  Rev.} \textbf{\bibinfo{volume}{251}}, \bibinfo{pages}{1007}
  (\bibinfo{year}{2007}).

\bibitem[{\citenamefont{Lopez et~al.}(2015)\citenamefont{Lopez, Piris,
  Ruip{\'{e}}rez, and Ugalde}}]{Lopez2015}
\bibinfo{author}{\bibfnamefont{X.}~\bibnamefont{Lopez}},
  \bibinfo{author}{\bibfnamefont{M.}~\bibnamefont{Piris}},
  \bibinfo{author}{\bibfnamefont{F.}~\bibnamefont{Ruip{\'{e}}rez}},
  \bibnamefont{and} \bibinfo{author}{\bibfnamefont{J.~M.}
  \bibnamefont{Ugalde}}, \bibinfo{journal}{J. Phys. Chem. A}
  \textbf{\bibinfo{volume}{119}}, \bibinfo{pages}{6981} (\bibinfo{year}{2015}).

\bibitem[{\citenamefont{Basch and Hoz}(1997)}]{Basch1997}
\bibinfo{author}{\bibfnamefont{H.}~\bibnamefont{Basch}} \bibnamefont{and}
  \bibinfo{author}{\bibfnamefont{S.}~\bibnamefont{Hoz}}, \bibinfo{journal}{J.
  Phys. Chem. A} \textbf{\bibinfo{volume}{101}}, \bibinfo{pages}{4416}
  (\bibinfo{year}{1997}).

\bibitem[{\citenamefont{Coote}(2004)}]{Coote2004}
\bibinfo{author}{\bibfnamefont{M.}~\bibnamefont{Coote}}, \bibinfo{journal}{J.
  Phys. Chem. A} \textbf{\bibinfo{volume}{108}}, \bibinfo{pages}{3865}
  (\bibinfo{year}{2004}).

\bibitem[{\citenamefont{Temelso et~al.}(2006)\citenamefont{Temelso, Sherrill,
  Merkle, and Freitas}}]{Temelso2006}
\bibinfo{author}{\bibfnamefont{B.}~\bibnamefont{Temelso}},
  \bibinfo{author}{\bibfnamefont{C.~D.} \bibnamefont{Sherrill}},
  \bibinfo{author}{\bibfnamefont{R.~C.} \bibnamefont{Merkle}},
  \bibnamefont{and} \bibinfo{author}{\bibfnamefont{R.~A.}
  \bibnamefont{Freitas}}, \bibinfo{journal}{J. Phys. Chem. A}
  \textbf{\bibinfo{volume}{110}}, \bibinfo{pages}{11160}
  (\bibinfo{year}{2006}).

\bibitem[{\citenamefont{Vandeputte et~al.}(2007)\citenamefont{Vandeputte,
  Sabbe, Reyniers, {Van Speybroeck}, Waroquier, and Marin}}]{Vandeputte2007}
\bibinfo{author}{\bibfnamefont{A.~G.} \bibnamefont{Vandeputte}},
  \bibinfo{author}{\bibfnamefont{M.~K.} \bibnamefont{Sabbe}},
  \bibinfo{author}{\bibfnamefont{M.-F.} \bibnamefont{Reyniers}},
  \bibinfo{author}{\bibfnamefont{V.}~\bibnamefont{{Van Speybroeck}}},
  \bibinfo{author}{\bibfnamefont{M.}~\bibnamefont{Waroquier}},
  \bibnamefont{and} \bibinfo{author}{\bibfnamefont{G.~B.} \bibnamefont{Marin}},
  \bibinfo{journal}{J. Phys. Chem. A} \textbf{\bibinfo{volume}{111}},
  \bibinfo{pages}{11771} (\bibinfo{year}{2007}).

\bibitem[{\citenamefont{Shao et~al.}(2003)\citenamefont{Shao, Head-Gordon, and
  Krylov}}]{Shao2003}
\bibinfo{author}{\bibfnamefont{Y.}~\bibnamefont{Shao}},
  \bibinfo{author}{\bibfnamefont{M.}~\bibnamefont{Head-Gordon}},
  \bibnamefont{and} \bibinfo{author}{\bibfnamefont{A.~I.}
  \bibnamefont{Krylov}}, \bibinfo{journal}{The Journal of Chemical Physics}
  \textbf{\bibinfo{volume}{118}}, \bibinfo{pages}{4807} (\bibinfo{year}{2003}).

\bibitem[{\citenamefont{Krylov}(2006)}]{Krylov2006a}
\bibinfo{author}{\bibfnamefont{A.~I.} \bibnamefont{Krylov}},
  \bibinfo{journal}{Accounts of chemical research}
  \textbf{\bibinfo{volume}{39}}, \bibinfo{pages}{83} (\bibinfo{year}{2006}).

\bibitem[{\citenamefont{Ess and Cook}(2001)}]{Ess2012}
\bibinfo{author}{\bibfnamefont{D.}~\bibnamefont{Ess}} \bibnamefont{and}
  \bibinfo{author}{\bibfnamefont{T.}~\bibnamefont{Cook}}, \bibinfo{journal}{J.
  Phys. Chem. A} \textbf{\bibinfo{volume}{116}}, \bibinfo{pages}{4922}
  (\bibinfo{year}{2001}).

\bibitem[{\citenamefont{Menon et~al.}(2007)\citenamefont{Menon, Wood, Moran,
  and Radom}}]{Menon2007}
\bibinfo{author}{\bibfnamefont{A.~S.} \bibnamefont{Menon}},
  \bibinfo{author}{\bibfnamefont{G.~P.~F.} \bibnamefont{Wood}},
  \bibinfo{author}{\bibfnamefont{D.}~\bibnamefont{Moran}}, \bibnamefont{and}
  \bibinfo{author}{\bibfnamefont{L.}~\bibnamefont{Radom}}, \bibinfo{journal}{J.
  Phys. Chem. A} \textbf{\bibinfo{volume}{111}}, \bibinfo{pages}{13638}
  (\bibinfo{year}{2007}).

\bibitem[{\citenamefont{Menon and Radom}(2008)}]{Menon2008}
\bibinfo{author}{\bibfnamefont{A.~S.} \bibnamefont{Menon}} \bibnamefont{and}
  \bibinfo{author}{\bibfnamefont{L.}~\bibnamefont{Radom}}, \bibinfo{journal}{J.
  Phys. Chem. A} \textbf{\bibinfo{volume}{112}}, \bibinfo{pages}{13225}
  (\bibinfo{year}{2008}).

\bibitem[{\citenamefont{Lai et~al.}(2012)\citenamefont{Lai, Li, Chen, and
  Shaik}}]{Lai2012}
\bibinfo{author}{\bibfnamefont{W.}~\bibnamefont{Lai}},
  \bibinfo{author}{\bibfnamefont{C.}~\bibnamefont{Li}},
  \bibinfo{author}{\bibfnamefont{H.}~\bibnamefont{Chen}}, \bibnamefont{and}
  \bibinfo{author}{\bibfnamefont{S.}~\bibnamefont{Shaik}},
  \bibinfo{journal}{Angewandte Chemie (International ed. in English)}
  \textbf{\bibinfo{volume}{51}}, \bibinfo{pages}{5556} (\bibinfo{year}{2012}).

\bibitem[{\citenamefont{Carstensen et~al.}(2007)\citenamefont{Carstensen, Dean,
  and Deutschmann}}]{Carstensen2007}
\bibinfo{author}{\bibfnamefont{H.-H.} \bibnamefont{Carstensen}},
  \bibinfo{author}{\bibfnamefont{A.~M.} \bibnamefont{Dean}}, \bibnamefont{and}
  \bibinfo{author}{\bibfnamefont{O.}~\bibnamefont{Deutschmann}},
  \bibinfo{journal}{Proceedings of the Combustion Institute}
  \textbf{\bibinfo{volume}{31}}, \bibinfo{pages}{149} (\bibinfo{year}{2007}).

\bibitem[{\citenamefont{Huynh et~al.}(2008)\citenamefont{Huynh, Barriger, and
  Violi}}]{Huynh2008}
\bibinfo{author}{\bibfnamefont{L.~K.} \bibnamefont{Huynh}},
  \bibinfo{author}{\bibfnamefont{K.}~\bibnamefont{Barriger}}, \bibnamefont{and}
  \bibinfo{author}{\bibfnamefont{A.}~\bibnamefont{Violi}}, \bibinfo{journal}{J.
  Phys. Chem. A} \textbf{\bibinfo{volume}{112}}, \bibinfo{pages}{1436}
  (\bibinfo{year}{2008}).

\bibitem[{\citenamefont{Huynh et~al.}(2010)\citenamefont{Huynh, Carstensen, and
  Dean}}]{Huynh2010}
\bibinfo{author}{\bibfnamefont{L.~K.} \bibnamefont{Huynh}},
  \bibinfo{author}{\bibfnamefont{H.-h.} \bibnamefont{Carstensen}},
  \bibnamefont{and} \bibinfo{author}{\bibfnamefont{A.~M.} \bibnamefont{Dean}},
  \bibinfo{journal}{J. Phys. Chem. A} \textbf{\bibinfo{volume}{114}},
  \bibinfo{pages}{6594} (\bibinfo{year}{2010}).

\bibitem[{\citenamefont{Stadtman and Levine}(2003)}]{Stadtman2003}
\bibinfo{author}{\bibfnamefont{E.~R.} \bibnamefont{Stadtman}} \bibnamefont{and}
  \bibinfo{author}{\bibfnamefont{R.~L.} \bibnamefont{Levine}},
  \bibinfo{journal}{Amino Acids} \textbf{\bibinfo{volume}{25}},
  \bibinfo{pages}{207} (\bibinfo{year}{2003}).

\bibitem[{\citenamefont{Prousek}(2007)}]{Prousek2007}
\bibinfo{author}{\bibfnamefont{J.}~\bibnamefont{Prousek}},
  \bibinfo{journal}{Pure and Applied Chemistry} \textbf{\bibinfo{volume}{79}},
  \bibinfo{pages}{2325} (\bibinfo{year}{2007}).

\bibitem[{\citenamefont{Balasubramanian
  et~al.}(1998)\citenamefont{Balasubramanian, Pogozelski, and
  Tullius}}]{Balasubramanian1998}
\bibinfo{author}{\bibfnamefont{B.}~\bibnamefont{Balasubramanian}},
  \bibinfo{author}{\bibfnamefont{W.~K.} \bibnamefont{Pogozelski}},
  \bibnamefont{and} \bibinfo{author}{\bibfnamefont{T.~D.}
  \bibnamefont{Tullius}}, \bibinfo{journal}{Proceedings of the National Academy
  of Sciences of the United States of America} \textbf{\bibinfo{volume}{95}},
  \bibinfo{pages}{9738} (\bibinfo{year}{1998}).

\bibitem[{\citenamefont{Vereecken and Peeters}(2001)}]{Vereecken2001}
\bibinfo{author}{\bibfnamefont{L.}~\bibnamefont{Vereecken}} \bibnamefont{and}
  \bibinfo{author}{\bibfnamefont{J.}~\bibnamefont{Peeters}},
  \bibinfo{journal}{Chem. Phys. Lett.} \textbf{\bibinfo{volume}{333}},
  \bibinfo{pages}{162} (\bibinfo{year}{2001}).

\bibitem[{\citenamefont{Blanksby and Ellison}(2003)}]{Blanksby2003}
\bibinfo{author}{\bibfnamefont{S.~J.} \bibnamefont{Blanksby}} \bibnamefont{and}
  \bibinfo{author}{\bibfnamefont{G.~B.} \bibnamefont{Ellison}},
  \bibinfo{journal}{Accounts of chemical research}
  \textbf{\bibinfo{volume}{36}}, \bibinfo{pages}{255} (\bibinfo{year}{2003}).

\bibitem[{\citenamefont{Piris}(2018{\natexlab{b}})}]{Piris2018a}
\bibinfo{author}{\bibfnamefont{M.}~\bibnamefont{Piris}}, in
  \emph{\bibinfo{booktitle}{Theoretical and Quantum Chemistry at the Dawn of
  the 21st Century}}, edited by
  \bibinfo{editor}{\bibfnamefont{T.}~\bibnamefont{Chakraborty}}
  \bibnamefont{and}
  \bibinfo{editor}{\bibfnamefont{R.}~\bibnamefont{Carb{\'{o}}-Dorca}}
  (\bibinfo{publisher}{Apple Academic Press}, \bibinfo{address}{New Jersey},
  \bibinfo{year}{2018}{\natexlab{b}}), chap.~\bibinfo{chapter}{22}, pp.
  \bibinfo{pages}{593--620}.

\bibitem[{\citenamefont{Piris and Ugalde}(2009)}]{Piris2009a}
\bibinfo{author}{\bibfnamefont{M.}~\bibnamefont{Piris}} \bibnamefont{and}
  \bibinfo{author}{\bibfnamefont{J.~M.} \bibnamefont{Ugalde}},
  \bibinfo{journal}{J. Comput. Chem.} \textbf{\bibinfo{volume}{30}},
  \bibinfo{pages}{2078} (\bibinfo{year}{2009}).

\bibitem[{\citenamefont{Zhao and Truhlar}(2008)}]{Zhao2008}
\bibinfo{author}{\bibfnamefont{Y.}~\bibnamefont{Zhao}} \bibnamefont{and}
  \bibinfo{author}{\bibfnamefont{D.~G.} \bibnamefont{Truhlar}},
  \bibinfo{journal}{Theoretical Chemistry Accounts}
  \textbf{\bibinfo{volume}{120}}, \bibinfo{pages}{215} (\bibinfo{year}{2008}).

\bibitem[{\citenamefont{Dunning and {Dunning Jr.}}(1989)}]{Dunning1989}
\bibinfo{author}{\bibfnamefont{T.~H.} \bibnamefont{Dunning}} \bibnamefont{and}
  \bibinfo{author}{\bibfnamefont{T.~H.} \bibnamefont{{Dunning Jr.}}},
  \bibinfo{journal}{J. Chem. Phys.} \textbf{\bibinfo{volume}{90}},
  \bibinfo{pages}{1007} (\bibinfo{year}{1989}).

\bibitem[{\citenamefont{{Johnson III}}(2018)}]{CCCBDB}
\bibinfo{editor}{\bibfnamefont{R.~D.} \bibnamefont{{Johnson III}}}, ed.,
  \emph{\bibinfo{title}{{NIST CCCBDB, NIST Standard Reference Database Number
  101, Release 19}}} (\bibinfo{year}{2018}).

\bibitem[{\citenamefont{Piris}(2014)}]{Piris2014c}
\bibinfo{author}{\bibfnamefont{M.}~\bibnamefont{Piris}}, \bibinfo{journal}{J.
  Chem. Phys.} \textbf{\bibinfo{volume}{141}}, \bibinfo{pages}{044107}
  (\bibinfo{year}{2014}).

\bibitem[{\citenamefont{Roos et~al.}(1980)\citenamefont{Roos, Taylor, and
  Siegbahn}}]{Roos1980}
\bibinfo{author}{\bibfnamefont{B.~O.} \bibnamefont{Roos}},
  \bibinfo{author}{\bibfnamefont{P.~R.} \bibnamefont{Taylor}},
  \bibnamefont{and} \bibinfo{author}{\bibfnamefont{P.~E.~M.}
  \bibnamefont{Siegbahn}}, \bibinfo{journal}{Chem. Phys.}
  \textbf{\bibinfo{volume}{48}}, \bibinfo{pages}{157} (\bibinfo{year}{1980}).

\bibitem[{\citenamefont{Siegbahn et~al.}(1980)\citenamefont{Siegbahn, Heiberg,
  Roos, and Levy}}]{Siegbahn1980}
\bibinfo{author}{\bibfnamefont{P.}~\bibnamefont{Siegbahn}},
  \bibinfo{author}{\bibfnamefont{A.}~\bibnamefont{Heiberg}},
  \bibinfo{author}{\bibfnamefont{B.~O.} \bibnamefont{Roos}}, \bibnamefont{and}
  \bibinfo{author}{\bibfnamefont{B.}~\bibnamefont{Levy}},
  \bibinfo{journal}{Phys. Scr.} \textbf{\bibinfo{volume}{21}},
  \bibinfo{pages}{323} (\bibinfo{year}{1980}).

\bibitem[{\citenamefont{Andersson et~al.}(1992)\citenamefont{Andersson,
  Malmqvist, and Roos}}]{Anderson1992}
\bibinfo{author}{\bibfnamefont{K.}~\bibnamefont{Andersson}},
  \bibinfo{author}{\bibfnamefont{P.}~\bibnamefont{Malmqvist}},
  \bibnamefont{and} \bibinfo{author}{\bibfnamefont{B.~O.} \bibnamefont{Roos}},
  \bibinfo{journal}{J. Chem. Phys.} \textbf{\bibinfo{volume}{96}},
  \bibinfo{pages}{1218} (\bibinfo{year}{1992}).

\bibitem[{\citenamefont{Aquilante et~al.}(2009)\citenamefont{Aquilante, Vico,
  Ferr{\'{e}}, Ghigo, Malmqvist, Neogr{\'{a}}dy, Pedersen, N{\'{a}}k, Reiher,
  Roos et~al.}}]{Aquilante2009}
\bibinfo{author}{\bibfnamefont{F.}~\bibnamefont{Aquilante}},
  \bibinfo{author}{\bibfnamefont{L.~D.~E.} \bibnamefont{Vico}},
  \bibinfo{author}{\bibfnamefont{N.}~\bibnamefont{Ferr{\'{e}}}},
  \bibinfo{author}{\bibfnamefont{G.}~\bibnamefont{Ghigo}},
  \bibinfo{author}{\bibfnamefont{P.-{\aa}.} \bibnamefont{Malmqvist}},
  \bibinfo{author}{\bibfnamefont{P.}~\bibnamefont{Neogr{\'{a}}dy}},
  \bibinfo{author}{\bibfnamefont{T.~B.} \bibnamefont{Pedersen}},
  \bibinfo{author}{\bibfnamefont{M.~P.} \bibnamefont{N{\'{a}}k}},
  \bibinfo{author}{\bibfnamefont{M.}~\bibnamefont{Reiher}},
  \bibinfo{author}{\bibfnamefont{B.~O.} \bibnamefont{Roos}},
  \bibnamefont{et~al.}, \bibinfo{journal}{J. Comp. Chem.}
  \textbf{\bibinfo{volume}{31}}, \bibinfo{pages}{224} (\bibinfo{year}{2009}).

\bibitem[{\citenamefont{Ervin and DeTuri}(2002)}]{Ervin2002}
\bibinfo{author}{\bibfnamefont{K.}~\bibnamefont{Ervin}} \bibnamefont{and}
  \bibinfo{author}{\bibfnamefont{V.}~\bibnamefont{DeTuri}},
  \bibinfo{journal}{The Journal of Physical Chemistry A}
  \textbf{\bibinfo{volume}{106}}, \bibinfo{pages}{9947} (\bibinfo{year}{2002}),
  \urlprefix\url{http://pubs.acs.org/doi/abs/10.1021/jp020594n}.

\bibitem[{\citenamefont{Henry et~al.}(2001)\citenamefont{Henry, Parkinson,
  Mayer, and Radom}}]{Henry2001}
\bibinfo{author}{\bibfnamefont{D.~J.} \bibnamefont{Henry}},
  \bibinfo{author}{\bibfnamefont{C.~J.} \bibnamefont{Parkinson}},
  \bibinfo{author}{\bibfnamefont{P.~M.} \bibnamefont{Mayer}}, \bibnamefont{and}
  \bibinfo{author}{\bibfnamefont{L.}~\bibnamefont{Radom}}, \bibinfo{journal}{J.
  Phys. Chem. A} \textbf{\bibinfo{volume}{105}}, \bibinfo{pages}{6750}
  (\bibinfo{year}{2001}).

\bibitem[{\citenamefont{Korchowiec}(2002)}]{Korchowiec2002}
\bibinfo{author}{\bibfnamefont{J.}~\bibnamefont{Korchowiec}},
  \bibinfo{journal}{J. Phys. Organic Chem.} \textbf{\bibinfo{volume}{15}},
  \bibinfo{pages}{524} (\bibinfo{year}{2002}).

\bibitem[{\citenamefont{Wood et~al.}(2005)\citenamefont{Wood, Moran, Jacob, and
  Radom}}]{Wood2005}
\bibinfo{author}{\bibfnamefont{G.~P.~F.} \bibnamefont{Wood}},
  \bibinfo{author}{\bibfnamefont{D.}~\bibnamefont{Moran}},
  \bibinfo{author}{\bibfnamefont{R.}~\bibnamefont{Jacob}}, \bibnamefont{and}
  \bibinfo{author}{\bibfnamefont{L.}~\bibnamefont{Radom}}, \bibinfo{journal}{J.
  Phys. Chem. A} \textbf{\bibinfo{volume}{109}}, \bibinfo{pages}{6318}
  (\bibinfo{year}{2005}).

\end{thebibliography}
\end{document}